\documentclass{elsarticle}
\usepackage{amssymb}
\usepackage{amsmath}
\usepackage{amsfonts}
\usepackage{enumerate}
\usepackage{graphicx}

\begin{document}

\title{Beyond the speed of light on Finsler spacetimes}

\author{Christian Pfeifer}
\ead{christian.pfeifer@desy.de}
\author{Mattias N.\,R. Wohlfarth}
\ead{mattias.wohlfarth@desy.de}

\address{II. Institut f\"ur Theoretische Physik und Zentrum f\"ur Mathematische Physik, Universtit\"at Hamburg, Luruper Chaussee 149, 22761 Hamburg, Germany}


\begin{abstract}
As a protoypical massive field theory we study the scalar field on the recently introduced Finsler spacetimes. We show that particle excitations exist that propagate faster than the speed of light recognized as the boundary velocity of observers. This effect  appears already in Finsler spacetime geometries with very small departures from Lorentzian metric geometry. It switches on for a sufficiently large ratio of the particle four-momentum and mass, and is the consequence of a modified version of the Coleman--Glashow velocity dispersion relation. The momentum dispersion relation on Finsler spacetimes is shown to be the same as on metric spacetimes, which differs from many quantum gravity models. If similar relations resulted for fermions on Finsler spacetimes, these generalized geometries could explain the potential observation of superluminal neutrinos claimed by the Opera collaboration.
\end{abstract}

\maketitle

It is generally held that particles cannot move faster than light. The deeper reason for this comes from classical field theory. There, particles are described by partial differential equations. The requirement that the Cauchy initial value problem be well-posed implies that the leading second order differential operator must be hyperbolic. If the geometric structure of the spacetime background is given solely by a Lorentzian metric, as in the Standard Model, hyperbolicity of the free field equations can only be related to the Lorentzian cone structure determined by this metric~\cite{Ratzel:2011zz}. It follows that the support of the fields must propagate in timelike directions~\cite{Benzoni} which in turn holds for the particle excitations of the field. Hence, Lorentzian geometry forces particle motion slower than light.

The study of deviations from Lorentzian geometry has a certain history in the context of quantum gravity models, see the review~\cite{AmelinoCamelia:2008qg}, and in that of geometric causality structures~\cite{Ratzel:2011zz}. These models often employ modified dispersion relations~\cite{moddisp}, some of which point towards a Finsler geometric origin~\cite{Finslerdispersion}. The questions arise whether these now consistently allow particles moving faster than light, and what is the underlying geometry? This is not only of theoretical interest, but could also bear on the reported observation of superluminal neutrinos by Opera~\cite{OPERA:2011zb}. Here we will investigate in some detail the consequences of an underlying Finsler spacetime geometry.

Finsler geometry realizes the weak equivalence principle through the most general geometric clock postulate for which proper time $T[x]$ depends on the position and four-velocity of observers or massive particles, moving along a worldline $x(\tau)$ through the spacetime manifold~$M$,
\begin{equation}\label{eq:Finsler}
T[x]=\int d\tau\,F(x(\tau),\dot x(\tau))\,.
\end{equation}
The metric limit is given by the tangent bundle function $F(x,y)=|\tilde g_{ab}(x)y^ay^b|^{1/2}$. Not all more general functions $F$ are suitable for physics. To see which are, we have developed the geometry of Finsler spacetimes in~\cite{Pfeifer:2011tk}. These provide the notions of null geometry and causal structure which are not available in standard mathematical settings of Finsler geometry, but are required physically to describe the propagation of light and define observers. 

In this letter we will exploit the controlled geometrical framework of Finsler spacetimes to analyze the propagation of the massive scalar field. Already on Finsler backgrounds mildly departing from Lorentzian geometry, we will demonstrate the existence of particle excitations that propagate faster than the speed of light recognized as the boundary velocity that observers cannot reach. These particle modes are characterized by large ratios of their four-momentum components and mass. This effect has a fully Finsler geometrical origin, and is the direct consequence of a more intricate velocity dispersion relation than on metric spacetimes. Since the scalar field is the prototypical massive field theory, a similar mechanism should be induced by Finsler spacetime geometry also on massive fermion fields. Experimental results on the observation of superluminal particles, e.g. by Opera, if confirmed, could then be explained; reversely, non-observations can be used to constrain Finsler deviations from Lorentzian geometry.

We will now make these statements precise. The theory presented here lives on the tangent bundle $TM$ of the spacetime manifold $M$ which is the union of all tangent spaces. In coordinates $(x^a)$ on some open neighbourhood $U\subset M$ one can write a vector $Y$ in $T_xM$ as $Y=y^a\partial_a{}_{|x}$; regarded as a point in $TM$, this vector has coordinates $(x^a,y^b)$. The associated coordinate basis of $TTM$ is denoted by $\{\partial_a, \bar\partial_a = \partial /\partial y^a\}$. We first review our definition of Finsler spacetimes, see~\cite{Pfeifer:2011tk} for details, and comment on the definition of observers and the measurement of velo\-cities, before we enter the discussion of the scalar field.

\vspace{3pt}\noindent\textit{Definition.} A Finsler spacetime $(M,L,F)$ is a four-dimensional smooth manifold~$M$ with a continuous function $L:TM\rightarrow\mathbb{R}$ on the tangent bundle which has the following properties: (a)~$L$ is smooth on $TM\setminus\{0\}$; (b)~$L$ is positively homogeneous of real degree $r\ge 2$ as $L(x,\lambda y)  = \lambda^r L(x,y)$ for all $\lambda>0$; (c) $L$ is reversible in the sense $|L(x,-y)|=|L(x,y)|$; (d) the Hessian $g^L_{ab}$ of $L$ with respect to the fibre coordinates is non-degenerate on $TM\setminus A$ where~$A$ has measure zero and does not contain the null structure $\{L(x,y)=0\}\subset TM$,
\begin{equation}
g^L_{ab}(x,y) = \frac{1}{2}\bar\partial_a\bar\partial_b L\,;
\end{equation}
(e) the unit timelike condition holds: for all $x\in M$,
\begin{eqnarray}
\Omega_x  =  \Big\{y\in T_xM \!\! &\Big|  & \!\! |L(x,y)|=1\,,\;g^L_{ab}(x,y)\textrm{ has}\nonumber\\
&& \textrm{signature }(\epsilon,-\epsilon,-\epsilon,-\epsilon)\,,\;\epsilon=\frac{|L(x,y)|}{L(x,y)}\Big\}
\end{eqnarray}
has a non-empty closed connected component $S_x$. The Finsler function $F(x,y) = |L(x,y)|^{1/r}$, the Finsler metric
\begin{equation}\label{eq:Fmet}
g^F_{ab}(x,y) = \frac{1}{2}\bar\partial_a\bar\partial_b F^2\,.
\end{equation}

\vspace{6pt}Finsler spacetimes generalize Lorentzian metric spacetimes $(M,\tilde g)$ with metric $\tilde g$ of signature $(-,+_3)$ and $L(x,y)=\tilde g_{ab}(x)y^ay^b$. This is homogeneous of order $r=2$; properties (a)--(c) are immediate; (d) holds since $g^L_{ab}=\tilde g_{ab}$ is invertible on $TM$; the unit timelike condition (e) is satisfied by the set $S_x$ of unit $\tilde g$-timelike vectors.
Finsler spacetimes vary from Finsler spaces defined in the literature, cf. the references in~\cite{Pfeifer:2011tk}. The central new ingredient is the function $L$ that acts as the fundamental geometric background structure. Properties (a)--(d) ensure that the geometry of the null structure $\{L=0\}$ is under control, and that null Finsler geodesics are well-defined. These are essential to describe the effective motion of massless particles and light. Condition~(e) implies the existence of  open convex cones of timelike vectors at all points $x$, constructed from the shell $S_x$ of unit timelike vectors; these have null boundary, and are required to model causality and four-velocities of physical observers. The Finsler function $F$ in~(\ref{eq:Finsler}) is a derived quantity.

Modelling an observer requires a frame $e_\mu=(e_0,e_\alpha)$ of tangent vectors $e_0$ and three orthogonal $e_\alpha$ which measure units of time and space, see~\cite{Pfeifer:2011xi} for details. When needed we identify tangent spaces $T_xM$ with horizontal tangent spaces $H_{(x,y)}TM$ by the isomorphism $Y^a\partial_a{}_{|x}\leftrightarrow Y^a\delta_a{}_{|(x,y)}$, where $\delta_a = \partial_a - N^b{}_a \bar\partial_b$ and the coefficients of the Cartan non-linear connection $N^a{}_b=\frac{1}{4}\bar\partial_b[g^{L\,ap}(y^q\partial_m\bar\partial_pL-\partial_pL)]$; in the metric limit these reduce to the Christoffel symbols $N^a{}_b\rightarrow\Gamma[\tilde g]^a{}_{bc}y^c$. The time direction of an observer with worldline $\gamma$ is the timelike normalized tangent~${e_0=\dot\gamma\in S_x}$; this implies $g^F_{(\gamma,e_0)}(e_0,e_0)=1$. Spatial directions are defined as the horizontal directions conormal to $dL$ which is equivalent to $g^F_{(\gamma,e_0)}(e_0,e_\alpha)=0$. The requirement that the $e_\alpha$ be orthogonal and measure unit length is modelled by $g^F_{(\gamma,e_0)}(e_\alpha,e_\beta)=-\delta_{\alpha\beta}$. Hence in the observer's frame
\begin{equation}\label{eq:onorm}
g^F_{(\gamma,e_0)}(e_\mu,e_\nu)=-\eta_{\mu\nu}
\end{equation}
with the metric limit $\tilde g_\gamma(e_\mu,e_\nu)=+\eta_{\mu\nu}$.

An observer measures the three-velocity components of a massive or null particle with worldline tangent $X=X^0e_0+\vec X=X^0e_0+X^\alpha e_\alpha$ as $v^\alpha = X^\alpha/X^0$, and 
\begin{equation}\label{eq:v2}
v_{(\gamma,e_0)}^2=\delta_{\alpha\beta}v^\alpha v^\beta = -\frac{g^F_{(\gamma,e_0)}(\vec X,\vec X)}{g^F_{(\gamma,e_0)}(e_0,X)^2}\,.
\end{equation}
Below we compare the velocity of massive scalar particle modes to the speed of light describing  the boundary velocity of observers. This is obtained from the maximal positive solution $X^0(\vec X)$  of the null condition $L(\gamma,X)=L(\gamma,X^0 e_0 + \vec X)=0$. Taylor expanding this around $\vec X=\vec 0$ and using $g^L_{(\gamma,e_0)}(e_\alpha,e_\beta) = \frac{\epsilon r}{2} g^F_{(\gamma,e_0)}(e_\alpha,e_\beta)$, we find
\begin{equation}\label{eq:velbound}
c_{(\gamma,e_0)}^2 = \frac{2}{r}+ \frac{2\epsilon}{r} \sum_{k=3}^\infty \frac{(X^0)^{-k}}{k!}\bar\partial_{c_1}\!\dots \bar\partial_{c_k}L_{(\gamma,e_0)}\vec X^{c_1}\!\dots \vec X^{c_k}\,.
\end{equation}

To learn more about the velocities of massive scalar particles, we study the relevant field theory. Action principles for fields on Finsler spacetime can be obtained from the corresponding action integral over the spacetime manifold. The procedure was tested in~\cite{Pfeifer:2011tk} for electrodynamics, where we could prove that light propagates along Finsler null geodesics. For scalar fields we start with the standard Lagrangian on metric spacetime~$(M,\tilde g)$,
\begin{equation}
\mathcal{L}[\tilde g,\tilde\phi,\partial\tilde\phi] = -\frac{1}{2} \tilde g^{ab}(x)\partial_a\tilde\phi(x)\partial_b\tilde\phi(x)-\frac{1}{2}m^2\tilde\phi(x)^2\,.
\end{equation}
Note that $\mathcal{L}[\dots]$ can be seen as a prescription to form a scalar quantity from tensorial objects. The same prescription generates a Lagrangian on $TM$, after promoting $\tilde \phi(x)$ to a tangent bundle field $\phi(x,y)$ of zero homogeneity in $y$, exchanging the partial derivatives on $M$ for partial derivatives on $TM$, and replacing the metric $\tilde g(x)$ by the following Sasaki-type metric, where $\delta y^a=dy^a+N^a{}_bdx^b$, 
\begin{equation}\label{eq:metric}
G_{(x,y)} = - g^F_{ab}(x,y) dx^a \otimes dx^b - \frac{g^F_{ab}(x,y)}{|L(x,y)|^{2/r}}\delta y^a\otimes \delta y^b\,.
\end{equation}

Action integrals on Finsler spacetimes are formulated on the seven-dimensional subbundle $\Sigma\subset TM$ defined by $|L(x,y)|=1$ and non-degenerate $g^L_{(x,y)}$. Convenient coordinates $(\hat x^a,u^\alpha)$ on $\Sigma$ are constructed in~\cite{Pfeifer:2011tk}; a volume form is induced by the pullback $G^*$ of $G$,
\begin{equation}\label{eq:Gstar}
G^*=-g^F_{ab}{}_{|\Sigma}d\hat x^a \otimes d\hat x^b - (g^F_{ab}\partial_\alpha y^a\partial_\beta y^b)_{|\Sigma}\delta u^\alpha \otimes\delta u^\beta\,,
\end{equation}
where $\delta u^\alpha = du^\alpha + \tilde N^\alpha{}_a d\hat x^a$ and $\tilde N^\alpha{}_a = N^p{}_a \bar\partial_p u^\alpha - \partial_a u^\alpha$. Then the Finsler spacetime action for massive scalars is
\begin{eqnarray}
S[\phi] &=& \int_\Sigma d^4\hat xd^3u\sqrt{G^*}\, \mathcal{L}[G,\phi,\partial\phi]_{|\Sigma}\\
&=& -\frac{1}{2} \int_\Sigma d^4\hat xd^3u\sqrt{G^*} \Big[G^{AB}\partial_A\phi\partial_B\phi+m^2\phi^2\Big]_{|\Sigma}\,,\nonumber
\end{eqnarray}
where the capital indices $A,B$ label the eight induced coordinates $(x^a,y^b)$ on $TM$. We will see from the corresponding equations of motion below, that this field theory is designed to  reduce to standard massive scalar field theory on $M$ in the limit of metric geometry, i.e., for $L(x,y)=\tilde g_{ab}(x)y^ay^b$ and $\phi(x,y)=\tilde\phi(x)$.

The equations of motion for $\phi$ are obtained by variation. We first expand the action in the horizontal/vertical basis $\{\delta_a,\,\bar\partial_a\}$ of $TTM$, where~(\ref{eq:metric}) can be used, 
\begin{equation}
S[\phi] = \frac{1}{2} \int_\Sigma \sqrt{G^*}\,\Big[g^{F\,ab}\delta_a\phi\delta_b\phi + g^{F\,ab}\bar\partial_a\phi\bar\partial_b\phi - m^2\phi^2\Big]_{|\Sigma}\,.
\end{equation}
To find $\delta S[\phi]$ we use  integration by parts formulae that can be proven with the coordinate transformations detailed in~\cite{Pfeifer:2011tk}. For $n$-homogeneous $A^a(x,y)$ we have
\begin{eqnarray}
0 &=& \int_\Sigma \sqrt{G^*}\,\Big[\delta_a A^a + \big(\Gamma^{\delta\,p}{}_{pa}+S^p{}_{pa}\big)A^a\Big]_{|\Sigma}\,,\\
0 &=& \int_\Sigma \sqrt{G^*}\,\Big[\bar\partial_a A^a + \big(g^{F\,pq}\bar\partial_ag^F_{pq} - (n+3) y^p g^F_{pa}\big)A^a\Big]_{|\Sigma}\,,\nonumber
\end{eqnarray}
using the shorthand notation $\Gamma^{\delta\,a}{}_{bc} = \frac{1}{2} g^{F\,ap}(\delta_b g^F_{pc}+\delta_c g^F_{pb}-\delta_p g^F_{bc})$ and $S^a{}_{bc} = \Gamma^{\delta\,a}{}_{bc} - \bar\partial_b N^a{}_c$. With these technical preparations we find the equations of motion
\begin{equation}\label{eq:phi}
\Big[-g^{F\,ab}\Big(\delta_a\delta_b-\Gamma^{\delta\,p}{}_{ab}\delta_p+\bar\partial_a\bar\partial_b+S^p{}_{pa}\delta_b\Big)\phi-m^2\phi\Big]_{|\Sigma}=0\,.
\end{equation}
In metric geometry this reduces to the Klein--Gordon equation $(\tilde g^{ab}\nabla[\tilde g]_a\partial_b -m^2)\tilde\phi =0$ with metric $\tilde g$. 

We now study the propagation of $\phi$ using techniques from the analysis of partial differential equations~\cite{Benzoni}. In coordinates $(x^{\hat M})=(\hat x^a,u^\alpha)$ of $\Sigma$ with associated basis $\{\partial_{\hat M}\}=\{\hat\partial_a,\partial_\alpha\}$ of $T\Sigma$ we find the principal symbol 
\begin{equation}
-g^{F\,ab}{}_{|\Sigma}\Big[\hat\partial_a\hat\partial_b\phi - 2 \tilde N^\alpha{}_a \partial_\alpha \hat\partial_b \phi 
+ \big(\tilde N^\alpha{}_a \tilde N^\beta{}_b + \bar\partial_a u^\alpha \bar\partial_b u^\beta\big)\partial_\alpha\partial_\beta \phi \Big]_{|\Sigma}\,,
\end{equation}
which can be rewritten as $G^{*\,\hat M \hat N}\partial_{\hat M}\partial_{\hat N}\phi\,$. Below we will consider Finsler spacetimes that slightly depart from Lorentzian geometry and have two almost identical lightcones. Then we shall confirm that the pullback metric $G^*$ on $\Sigma$ is of Lorentzian signature $(-,+_6)$. This implies that the field equations of the massive scalar field~$\phi$ are hyperbolic, and so have a well-posed Cauchy problem. Allowed Cauchy surfaces of initial data are conormal to the momenta $P\in T^*\Sigma$ in the lightcone $G^{*\,-1}(P,P)<0$. Moreover, the support of the field $\phi$ on $\Sigma$ propagates into timelike directions $X$ in the lightcone of $G^*$  in $T\Sigma$. These are obtained from the momenta by the map 
\begin{equation}\label{eq:mv}
X=\frac{1}{m}G^{*\,-1}(P,\cdot)\,.
\end{equation}

Only those excitations of $\phi$ moving tangent to the spacetime $M$ can be interpreted as classical particles. The relevant tangents are identified with the  horizontal vector fields over $\Sigma\subset TM$ and can be expressed as $X=X^a\hat\delta_a=X^a(\hat \partial_a- \tilde N^\alpha{}_a \partial_\alpha)$. Using~(\ref{eq:mv}) and the fact that $G^*$ preserves the horizontal and vertical structure according to~(\ref{eq:Gstar}), shows that also particle momenta must be horizontal, i.e., $P=P_a d\hat x^a$.

\begin{figure}[ht]
\begin{center}
\includegraphics[angle=270,width=0.6\columnwidth]{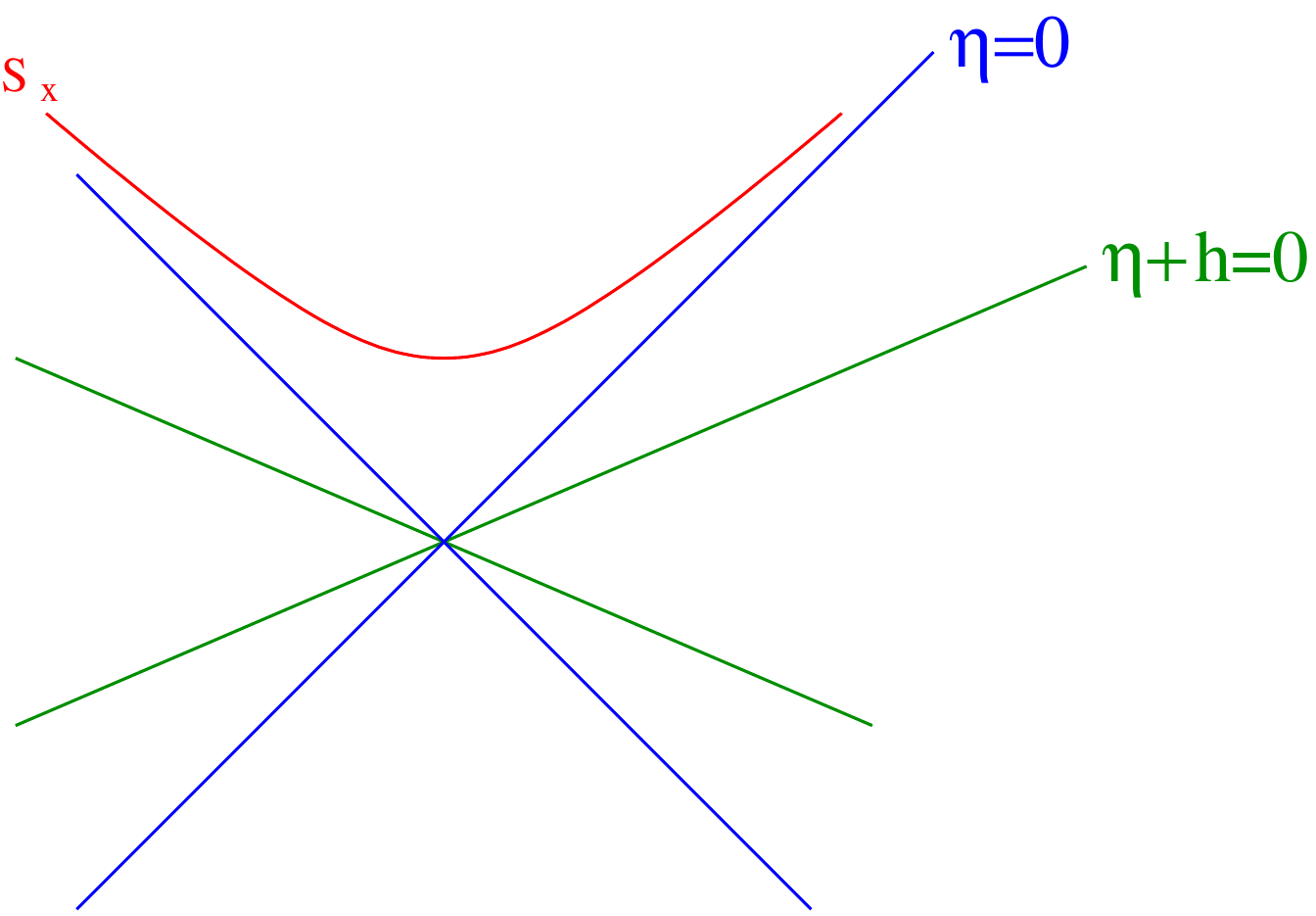}
\caption{\label{fig:1}Null structure and unit timelike shell in $T_xM$ at any point $x$ of a simple bimetric Finsler spacetime.}
\end{center}
\end{figure}

Now we are in the position to study the dispersion relation of the particle excitations of~$\phi$. We specify a Finsler spacetime mildly departing from flat Lorentzian metric geometry, where standard dispersion relations are well-defined. We consider the simple bimetric background
\begin{equation}\label{eq:L}
L(x,y) = \eta_{ab}y^ay^b\left(\eta_{cd}+h_{cd}\right)y^cy^d
\end{equation}
without $x$-dependence. $\eta_{ab}$ denotes the usual Minkowski metric, and $h_{cd}$ is chosen so that $\eta_{ab}+h_{ab}$ is Lorentzian and has a timelike cone containing that of $\eta_{ab}$. The null structure of $L$ is the union of the light cones of $\eta$ and $\eta+h$. The function $L$ defines a Finsler spacetime with homogeneity $r=4$; for small components $h_{ab}$, one can check that the closed connected component $S_x$ of unit timelike vectors is given by $\eta(y,y)=-1$ up to a perturbation; hence observers move along $\eta$-timelike worldlines. These characteristics are shown in figure~\ref{fig:1}.

The dispersion relation is derived from the field equation~(\ref{eq:phi}). This simplifies on the Finsler spacetime defined by $L$ in~(\ref{eq:L}), where both $N^a{}_b=0$ and $\Gamma^{\delta\,a}{}_{bc}=0$, to
\begin{equation}
-g^{F\,ab}{}_{|\Sigma}\Big(\hat\delta_a\hat\delta_b+\bar\partial_au^\alpha\bar\partial_bu^\beta \partial_\alpha\partial_\beta+\bar\partial_a\bar\partial_bu^\alpha\partial_\alpha
\Big)_{|\Sigma}\phi=m^2\phi\,.
\end{equation} 
In a Fourier decomposition into modes of momentum $P=P_a d\hat x^a+P_\alpha\delta u^\alpha=(P_a+\tilde N^\alpha{}_aP_\alpha)d\hat x^a+P_\alpha du^\alpha$ of the form $\exp[i(P_a+\tilde N^\alpha{}_aP_\alpha)\hat x^a+iP_\alpha u^\alpha]$ the operators $\hat\delta_a$ and $\partial_\alpha$ act as multiplication by $iP_a$ and $iP_\alpha$. Particle modes have horizontal momentum, so $P_\alpha=0$. For the particle excitations of the massive scalar field we hence find the dispersion relation $-g^{F\,ab}P_aP_b=-m^2$ . In any observer frame with~(\ref{eq:onorm}) this has the standard form for the measurable components $P_\mu$ of 4-momentum,
\begin{equation}\label{eq:disp}
-g^F_{(\gamma,e_0)}(P,P) = \eta^{\mu\nu}P_\mu P_\nu = -m^2\,.
\end{equation}

We now interpret this result. The Finsler metric~(\ref{eq:Fmet}) of the bimetric geometry~(\ref{eq:L})  is $-g^F_{ab}(\gamma,e_0)= \eta_{ab}+h_{ab}/2+h^{(2)}_{ab}(e_0)$ to second order in the components $h_{ab}$. So the signature of $-g^F_{ab}$ is $(-,+_3)$; comparing (\ref{eq:metric}) and (\ref{eq:Gstar}) similarly as in~\cite{Pfeifer:2011tk} then shows $G^*$ has the Lorentzian signature $(-,+_6)$ required for a well-posed propagation of~$\phi$. 
Qualitative effects of the generalized spacetime geometry already appear at first order in $h_{ab}$. Substituting $g^F_{ab}$ into the dispersion relation, and using~(\ref{eq:mv}) yields
\begin{equation}\label{eq:disp2}
\eta_{ab}X^aX^b = -1-\frac{h^{ab}P_aP_b}{2\,m^2} + \mathcal{O}(h^2)
\end{equation}
where $h^{ab}=\eta^{ap}\eta^{bq}h_{pq}$, and $X^a$ and $P_a$ are the horizontal components of particle velocity and momentum. The key feature of the formula above is that massive particles are not always $\eta$-timelike as are the observers. If the ratio $h^{ab}P_aP_b/m^2<0$ and if $P_a/m$ are sufficiently large, then the massive particles are faster than the boundary velocity for observers given by light on $\{\eta(y,y)=0\}$. We remark that the massive scalar field has particle excitations moving in $\eta$-timelike, $\eta$-null and $\eta$-spacelike directions. 

To find the quantitative difference between the massive particles' velocity and the boundary velocity of observers we use the second order expansion of $g^F$. The series~(\ref{eq:velbound}) terminates at $k=4$; to calculate the required maximal positive solution $X^0(\vec X)$ of the null condition $L(\gamma,X)=0$ we rewrite this condition in the observer's frame as
\begin{equation}
 (X^0)^4-2(X^0)^2\vec X^2 + (\vec X^2)^2 + X^0 C_3(\vec X) +C_4(\vec X) =\mathcal{O}(h^3) \,.
\end{equation}
Then $X^0(\vec X) = \sqrt{\vec X^2} \big( 1 + M_{(\gamma,e_0)}(\vec X)/2\big)+\mathcal{O}(h^2)$ where
\begin{equation}
M_{(\gamma,e_0)}(\vec X) = \frac{(-\sqrt{\vec X^2}C_3(\vec X)-C_4(\vec X))^{1/2}}{\vec X^2}
\end{equation}
with $-3C_3(\vec X)=\bar\partial_a h^{(2)}_{bc}(\gamma,e_0)\vec X^a\vec X^b\vec X^c$ and $-6C_4(\vec X)=\bar\partial_a \bar\partial_b h^{(2)}_{cd}(\gamma,e_0)\vec X^a ... \vec X^d$. For any observer exist four solutions of the null condition, hence $M_{(\gamma,e_0)}(\vec X)$ is positive real. With this result we find the boundary velocity
\begin{equation}
c_{(\gamma,e_0)}^2 = 1- M_{(\gamma,e_0)}(\vec X) +\mathcal{O}(h^2)\,.
\end{equation}
Finally, (\ref{eq:v2}) and (\ref{eq:disp}) yield the velocity dispersion relation
\begin{equation}\label{eq:veldiff}
v_{(\gamma,e_0)}^2 - c_{(\gamma,e_0)}^2 = -\frac{m^2}{P_0^2} + M_{(\gamma,e_0)}(\vec X) +\mathcal{O}(h^2) \,.
\end{equation}
$M_{(\gamma,e_0)}(\vec X)$ is positive for any given spacetime, observer $(\gamma,e_0)$, and spatial direction of motion $\vec X$ of the massive particle and the reference light ray; this dependence modifies the velocity dispersion relation in~\cite{AmelinoCamelia:2011dx} with constant $M$ which is inspired from the result derived by 
Coleman--Glashow in a Lorentz symmetry violating extension of the Standard Model~\cite{Coleman:1998ti}. Here, superluminal particle modes occur, if $P_0/m$ is sufficiently large.  Moreover, any given spacetime solution of the Finsler gravity equation, see \cite{Pfeifer:2011xi}, immediately provides $M_{(\gamma,e_0)}(\vec X)$ explicitly in terms of the solution's parameters. In this way, bounds on these gravitational parameters can be found from particle physics experiments, or, if the parameters were already determined by gravity experiments, one could predict the attainable superluminality of massive particles.

\vspace{3pt}\noindent\textit{Discussion.} We have studied the propagation of the scalar field on Finsler spacetimes. As central results we find in any observer's frame the standard momentum dispersion relation~(\ref{eq:disp}) for the massive field modes, while the corresponding velocity dispersion relation~(\ref{eq:veldiff}) is modified due to Finsler geometric effects. In consequence, superluminal particle modes of the massive scalar field can be measured when the ratio $P_0/m$ becomes sufficiently large. As an inherently geometric effect, this result is expected for other massive, also fermionic, fields. The key feature we employed to derive the momentum and velocity dispersion relations is our well-defined notion of observers on a Finsler spacetime background; this gives us control over their notions of energy and velocity.

Since the momentum dispersion relation~(\ref{eq:disp}) is unchanged from the Lorentz\-ian one, it would be interesting to study whether there exist Finsler metrics that transform the velocity dispersion relations discussed e.g. in~\cite{AmelinoCamelia:2011dx} into the standard momentum dispersion relation for observers. These relations then can be explained in terms of a Finsler spacetime geometry. This would lead to a  geometric picture of modified velocity dispersion relations inspired from quantum gravity models or from Coleman--Glashow models. 

Can our result of superluminal massive particle modes on Finsler spacetimes be consistent with existing experimental constraints on the departure from Lorentzian geometry?

The non-observation of superluminal particles up to now, with the possible exception of the Opera results, immediately shows that the values of $M$ in~(\ref{eq:veldiff}) must be very small, since $M < m^2/P_0^2$ for all particle masses and experimentally reached energies. Since the function $M$ is determined by third and fourth $y$-derivatives of the Finsler function which are zero in the metric limit, this merely constrains the deviation from metric geometry. Note that the constraint obtained on $M$ could be very different in laboratory experiments and in astronomical observations, simply because the spherically symmetric Finsler spacetime in the vicinity of the Earth would be expected to  produce effects of different size as the cosmological spacetime relevant for observations in our Universe.
Possible future observations of superluminality will give us a clearer idea of the size of $M$. While superluminality is governed by this size, the anisotropy of the speed of light (boundary speed of observers) is given by $\Delta M_{(\gamma,e_0)} = \max_{\vec X} M_{(\gamma,e_0)}(\vec X) - \min_{\vec X}M_{(\gamma,e_0)}(\vec X)$. This quantity could be much smaller than $M$, depending on the Finsler spacetime solution. Even if superluminality is observed, this mechanism could avoid the strong bounds on Lorentz symmetry violations from anisotropy measurements of the speed of light.

In the light of the possible observation of superluminal motion reported by Opera, Finsler spacetimes lead to an explanation that differs from many others~\cite{OPERAexplanations} since it is based on changing the standard geometry that underlies physics. Indeed, we argued that superluminal velocities require a modification of Lorentzian spacetime geometry to alter the metric principal symbols in the free field equations, unless one appeals to an effective modification so that the fields are always interacting~\cite{nuinteraction}. Furthermore,  on Finsler spacetimes we do not expect problems with bremsstrahlung effects as discussed in~\cite{Cohen:2011hx} due to the following facts: our geometric definition of observer frames yields the standard momentum dispersion relation~(\ref{eq:disp}); our theory has no preferred frames which is crucial according to~\cite{AmelinoCamelia:2011bz}; the calculation~\cite{Chang:2012gr} for a concrete Finsler geometric model demonstrates that neutrino bremsstrahlung is excluded.

We conclude that Finsler spacetimes should be regarded as an interesting generalization of Lorentzian spacetimes: they provide a background for physics that offers the possibillity of superluminal motion without conceptional or experimental contradictions. The gravity equation for Finsler spacetimes is presented in \cite{Pfeifer:2011xi} and the study of solutions of this equation which go beyond metric geometry is an ongoing research project.

\vspace{3pt}\noindent\textit{Acknowledgments.} 
We thank Rutger Boels, Claudio Dappiaggi, Andreas Degner, Klaus Fredenhagen, Manuel Hohmann and Felix Tennie, and the German Research Foundation for support through grant WO~1447/1-1.

\end{document}